\documentclass[9pt,shortpaper,twoside,web]{ieeecolor}
\usepackage{generic}
\usepackage{cite}
\usepackage{amsmath,amssymb,amsfonts}
\usepackage{algorithmic}
\usepackage{graphicx}
\usepackage{textcomp}
\usepackage{stackengine}
\usepackage{color}
\usepackage{stackengine}
\usepackage{svg}
\usepackage{multirow}
\usepackage{easyReview}
\usepackage{subfigure}
\usepackage{url}

\def\BibTeX{{\rm B\kern-.05em{\sc i\kern-.025em b}\kern-.08em
    T\kern-.1667em\lower.7ex\hbox{E}\kern-.125emX}}
\begin{document}
\title{On Smart Gaze based Annotation of Histopathology Images for  Training of Deep Convolutional Neural Networks}
\author{Komal Mariam$^1$, Osama Mohammed Afzal$^1$, Wajahat Hussain$^1$, Muhammad Umar Javed$^1$, \\Amber Kiyani$^2$, Nasir Rajpoot$^3$,  
Syed Ali Khurram$^4$ and Hassan Aqeel Khan$^{5,1}$
\thanks{Manuscript accepted at IEEE JBHI (Journal of Biomedical and Health Informatics). \copyright~2022 IEEE. Personal use of this material is permitted. Permission from IEEE must be obtained for all other uses, including reprinting/republishing this material for advertising or promotional purposes, collecting new collected works for resale or redistribution to servers or lists, or reuse of any copyrighted component of this work in other works.}
\thanks{$^1$Komal Mariam, Osama Mohammed Afzal, Wajahat Hussain and Muhammad Umar Javed are with the National University of Sciences and Technology (NUST), H-12 Islamabad, Pakistan (e-mail: kmariam.msee17seecs@seecs.edu.pk; osama.bscs17seecs@seecs.edu.pk; wajahat.hussain@seecs.edu.pk; mjaved.bscs18seecs@seecs.edu.pk). }
\thanks{$^2$Amber Kiyani is with Riphah International University, Islamabad, Pakistan (e-mail: amber.kiyani@riphah.edu.pk).}
\thanks{$^3$Nasir Rajpoot is with the Department
of Computer Science, University of Warwick, Coventry, CV4 7AL UK (email: n.m.rajpoot@warwick.ac.uk).}
\thanks{$^4$Syed Ali Khurram is with The School of Clinical Dentistry, University
of Sheffield, 19 Claremont Crescent, Sheffield, S10 2TA UK (e-mail:
s.a.khurram@sheffield.ac.uk).}
\thanks{$^5$Hassan Aqeel Khan is with the College of Computer Science and Engineering, University of Jeddah, Jeddah, Saudi Arabia. He is on sabbatical from NUST,  Islamabad (e-mail: hakhan@uj.edu.sa).}
\thanks{Corresponding Author: Hassan Aqeel Khan}}

\maketitle

\begin{abstract}
Unavailability of large training datasets is a bottleneck that needs to be overcome to realize the true potential of deep learning in histopathology applications. Although slide digitization via whole slide imaging scanners has increased the speed of data acquisition, labeling of virtual slides requires a substantial time investment from pathologists. Eye gaze annotations have the potential to speed up the slide labeling process. This work explores the viability and timing comparisons of eye gaze labeling compared to conventional manual labeling for training object detectors. Challenges associated with gaze based labeling and methods to refine the coarse data annotations for subsequent object detection are also discussed. Results demonstrate that gaze tracking based labeling can save valuable pathologist time and delivers good performance when employed for training a deep object detector. Using the task of localization of Keratin Pearls in cases of oral squamous cell carcinoma as a test case, we compare the performance gap between deep object detectors trained using hand-labelled and gaze-labelled data. 
On average, compared to `Bounding-box' based hand-labeling, gaze-labeling required $57.6\%$ less time per label and compared to `Freehand' labeling, gaze-labeling required on average $85\%$  less time per label. 
\end{abstract}


%\section{Introduction}
%\label{sec:introduction}
\section{Introduction}
\label{sec:Intro}
\IEEEPARstart{T}{he} field of pathology has witnessed some exciting developments over the past decade. The emergence of Whole Slide Imaging  (WSI) Scanners has paved the way for a completely digital workflow in both diagnostic and research laboratories. Simultaneously, in deep learning, the community now has a set of powerful tools that can capitalise on the troves of digital WSI  data being generated by the scanners. %and significantly enhance the throughput of pathology service providers.  
 However, a number of challenges still need to be overcome in order to realise a truly digital, high-throughput pathology infrastructure.
 Deep learning algorithms are generally \textit{data hungry} and require large volumes of labelled data for training. Unfortunately, manual annotation and labeling of high resolution WSIs is a laborious process that demands investment of valuable time and effort from specialist pathologists. A typical WSI consists of tens of thousands of pixels at multiple magnification levels; studies suggest that labeling a single WSI can (in extreme cases) take up to 360 minutes \cite{Lindman2019}. Consequently, smart labeling strategies, that can reduce the amount of time and effort required for labeling of WSI data, have the potential to enhance the performance of automated pathology solutions.

\begin{figure}
\centering
\includegraphics[width=\columnwidth]{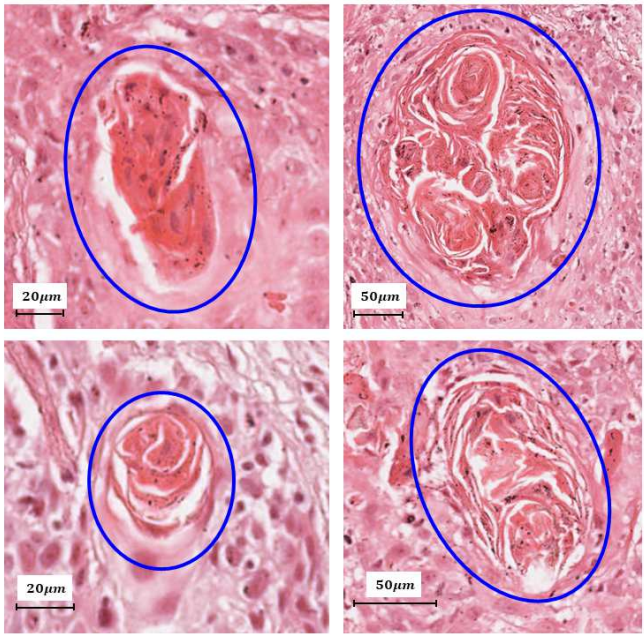}
  \caption{Examples of KPs of different shapes and sizes.}
  \label{fig:KP_Magnified}
\end{figure}

This work examines the feasibility of using gaze tracking for generation of large-scale, labelled training data from WSIs. More specifically, we employ low-cost eye-tracking hardware to record the gaze patterns of users as they explore a WSI on a desktop screen. We ask users to search for Keratin Pearls (KPs) in WSIs and investigate whether or not user gaze patterns can be used to train deep object detectors to localise KPs in unlabelled test data.
KP formation is one of the characteristic diagnostic features for well-differentiated oral squamous cell carcinoma. The pearls present as concentric masses of pink-colored keratinocytes and keratin protein that are easy to identify \cite{woolgar2009pitfalls, woolgar2006histopathological}. Four examples of KPs are shown in Fig.~\ref{fig:KP_Magnified}.
Only a few existing studies (see section \ref{sec:Related}) have examined the feasibility of using gaze tracking for labeling of medical imaging data and most of the reported results are preliminary. However, the following questions still remain unanswered:
\begin{itemize}
    \item[\emph{Q-1}] Compared to hand labeling, how much (if any) labeling noise is observed when gaze based labeling is employed?
    \item[\emph{Q-2}] Are there any pre-processing techniques that can reduce or eliminate label noise introduced by gaze based labeling?
    \item[\emph{Q-3}] How much (if any) time saving can gaze based labeling deliver?
    \item[\emph{Q-4}] Are there any limits on the size and shapes of objects/ features that can be labelled using gaze tracking?
    \item[\emph{Q-5}] Compared to training on hand labelled data, how much (if any) performance degradation occurs in well-established deep convolutional neural networks (CNNs) when they are trained on gaze annotated data? 
\end{itemize}

\begin{figure*}
\centering
\includegraphics[width=0.92\textwidth]{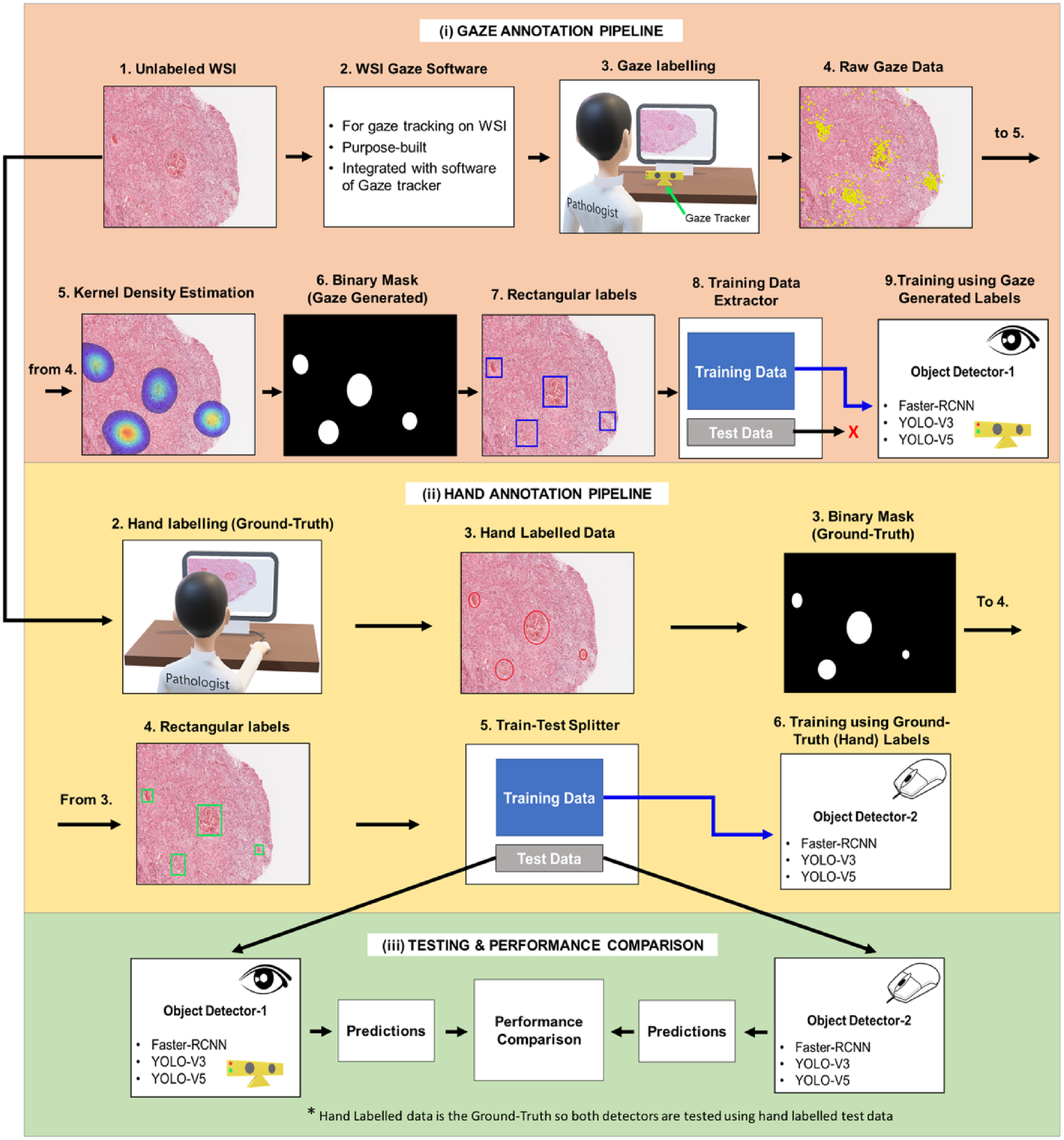}
  \caption{Gaze Annotation vs Hand Annotation Pipeline}
  \label{fig:pipeline}
\end{figure*}

Our objective in this work is to use an evidence based approach to answer the questions listed above. In order to obtain answers to these questions we first developed a purpose-built software which integrated our gaze tracking hardware with a WSI reader. Openslide \cite{goode2013openslide} was used to read WSIs in $.svs$ format. Our software can keep track of user gaze patterns across multiple magnification levels of a WSI. This tool was then employed to generate gaze labelled data using commercially available gaze tracking hardware. The resulting labelled data was then used to train deep object detectors to localize regions of interest (ROIs), and performance was compared with hand labelled data. We also present results of experiments that compare the amount of labeling time required by users when using conventional hand-labeling and gaze based labeling.

A high-level overview of our gaze annotation pipeline is shown in Fig.~\ref{fig:pipeline}. Our primary objective is to evaluate whether keeping track of the gaze patterns of pathologists can enable generation of useful labels that can be employed to train machine learning algorithms to detect ROIs in WSI data. To achieve this, a custom-built software was employed to track the gaze of skilled pathologists as they searched for target ROIs on WSIs on a computer screen. KP regions in 4 cases of oral squamous cell carcinoma (OSCC) were designated as the target ROIs. Preliminary analysis of raw gaze maps indicated that they contained large amounts of noise since pathologists were not looking at the target ROIs at all times during the annotation process. Then, Kernel Density Estimation (KDE) was employed to detect actual ROIs and eliminate noisy labels (details in section~\ref{subSec:KDE}). The output of the KDE block were heatmaps indicating the gaze patterns on the slide; these heatmaps were then converted into binary masks via thresholding and employed as training labels for an object detector. Three types of object detectors were tested in this study: Faster R-CNN \cite{ren2015faster}, YOLOv3 \cite{redmon2018yolov3} and YOLOv5 \cite{ultralytics}. For benchmarking, pathologists were asked to label KP regions on the same WSI data by hand, using a mouse and QuPath software\cite{Bankhead2017}. The resulting hand annotated label masks were employed to train the benchmark object detectors.
%The resulting, hand generated, label masks were employed to train, the Benchmark, Object Detector-2. 
For testing and performance comparison, we compared the predictions of Object Detector-1 (trained using gaze-labelled data) and Object Detector-2 (trained using hand-labelled data) with the ground-truth (hand annotated) labels corresponding to the unseen test dataset.
The amount of time required for annotation using both techniques was also recorded. It is highlighted that KP detection was selected as a use case primarily because, in general, KPs are neither too large nor too small in size and thus are relatively easier to detect. For this initial evaluation of gaze annotations we felt it was appropriate to limit ourselves to a labeling task which was neither too challenging nor too easy to perform. 
The main contributions of this work are listed below: 
\begin{itemize}
    \item We present the first of its kind, in-depth analysis on the suitability of using gaze based annotation of histopathology images for training deep CNNs.
    \item A novel dataset of gaze labelled histopathology images is released to the public for further experimentation and analysis. All data and code used for experiments in this work can be downloaded from the github repository available at: \url{https://github.com/SigmaLabResearch/Gaze-Enabled-Histopathology}
    \item A custom-built software tool that can use commercially available hardware to generate raw gaze maps on WSIs is also being released to the public. There are currently no open-source software tools that support gaze tracking on WSI file formats. This tool can be used to further enhance research in gaze based labeling of histopathology images and is available for download at: \url{https://github.com/SigmaLabResearch/Visnotate}.
    \item Results of a study comparing the amount of time saved by employing gaze based labeling are also presented.
\end{itemize}

\begin{table*}[]
\resizebox{\textwidth}{!}{%
{\renewcommand{\arraystretch}{1.4}
\begin{tabular}{|c|c|c|c|c|c|}
\hline
\textbf{Ref No.}                                                        & \textbf{Domain}         & \textbf{Annotation Time Provided?}                                                                          & \textbf{Gaze Constrained?} & \textbf{Approach}                                                         & \textbf{AI Based Performance Evaluation?}                                                                                     \\ \hline
Sadeghi \cite{Sadeghi2009}                                                             & Radiology + Dermatology & Yes                                                                                                         & Yes                       & Gaze based labels only                                                    & No                                                                                                                            \\ \hline
Stember \cite{Stember2019}                                                            & Radiology               & \begin{tabular}[c]{@{}c@{}}Partial (Time consumed for hand \\ labeling not provided)\end{tabular}           & Yes                       & Gaze based labels only                                                    & Yes   (UNet based segmentation)                                                                                               \\ \hline
\begin{tabular}[c]{@{}c@{}}Cai \cite{Cai2018}\\  (SonoEyenet)\end{tabular}         & Ultasound               & No                                                                                                          & No                        & \begin{tabular}[c]{@{}c@{}}Gaze + Visual \\ Attention Models \end{tabular} & Yes (CNN based detection)                                                                                                     \\ \hline
\begin{tabular}[c]{@{}c@{}}Khosravan \cite{Khosravan2017} \\ (Gaze2segment)\end{tabular} & Radiology               & No                                                                                                          & No                        & \begin{tabular}[c]{@{}c@{}}Gaze + Visual \\ Attention Models \end{tabular} & \begin{tabular}[c]{@{}c@{}}Yes (Random Walk \\ based segmentation)\end{tabular}                                               \\ \hline
Lejeune \cite{Lejeune2017}                                                             & Multiple                & No                                                                                                        &\begin{tabular}[c]{@{}c@{}} Yes \end{tabular}                        & Gaze based labels only                                                    & \begin{tabular}[c]{@{}c@{}}Yes (Semi-supervised classification \\ based on Expected Exponential Loss\\  function\end{tabular} \\ \hline
Vilarino \cite{Vilarino2007}                                                            & Colonoscopy             & No                                                                                                          & Yes                         & Gaze based labels only                                                     & Yes (SVM based classification)                                                                                                \\ \hline
Brunye \cite{Brunye2017}                                                           & WSI                     & No                                                                                                          & No                        & Gaze based labels only                                                    & No                                                                                                                            \\ \hline
Krupinski \cite{Krupinski2006}                                                           & WSI                     & \begin{tabular}[c]{@{}c@{}}Yes (Fixation time presented, \\ study did not focus on annotation)\end{tabular} & No                        & Gaze based labels only                                                    & No                                                                                                                            \\ \hline
Ersoy \cite{Ersoy2017}                                                             & WSI                     & No                                                                                                          & No                        & Gaze based labels only                                                    & No                                                                                                                            \\ \hline
Our work                                                                & WSI                     & Yes                                                                                                         & No                        & Gaze based labels only                                                    & \begin{tabular}[c]{@{}c@{}}Yes (CNN based object detection)\end{tabular}                                           \\ \hline
\end{tabular}%
}
}
  \caption{Summary of related work}
  \label{tab:RelatedWork}
\end{table*}

\section{Related Work}
\label{sec:Related}
Existing literature on gaze tracking primarily consists of approaches that have examined its impact on radiology and other (non-pathology) domains.  For example, in \cite{Sadeghi2009} gaze tracking was employed to label ROIs in radiology and skin lesion images. An 18.6\% improvement in labeling speed was observed, compared to conventional mouse-based hand labeling, in images with a high object-background contrast. The resulting labelled data was employed for segmentation of ROIs using a random walk based algorithm. However, the algorithm's performance was not quantified using any metrics. Furthermore, this approach was somewhat constraining since it required the labeling experts to deliberately shift focus from candidate ROIs to background pixels to ensure appropriate training for segmentation. 

Gaze labeling of radiology data was also examined in \cite{Stember2019} which employed UNet for segmentation of radiology images. Gaze and hand labelled images were observed to have an average Dice similarity coefficient of 0.85 with each other. For label generation, radiologists were required to explicitly move their eyes in a counterclockwise trace around an ROI's outer contour. Zoom/pan and window level adjustments were prohibited during contour tracing. A detailed comparison between the amount of time consumed by each labeling strategy was not provided. This approach also requires substantial post-processing effort for manual deletion of all gaze points recorded whilst the labeling expert is not looking at the target ROI, significantly constraining its use in practical applications.

Gaze patterns of sonographers were combined with image feature maps to measure abdominal circumference in ultrasound data in \cite{Cai2018}. Results demonstrated a marked improvement in performance compared to using only feature maps.
A similar approach was presented in \cite{Khosravan2017}; gaze patterns of radiologists were combined with computer generated saliency maps to segment ROIs using a random walk based segmentation algorithm. 

Other similar approaches on non-radiology data were presented in \cite{Lejeune2017} and \cite{Vilarino2007}, though these did not include thorough comparison with manual annotation in terms of labeling time. Another limitation of \cite{Lejeune2017} and \cite{Vilarino2007} is that only a single object or ROI was assumed to be present on screen at one time during labeling. In \cite{Lejeune2017}, an observer was asked to fixate on a target and it was assumed that observers were 100\% compliant and did not look at non-target regions at any time during recording.
%\textcolor{green}{\cite{Lejeune2017} only requires a single gaze position for each image and assumes the observer is complaint to the task. 
In \cite{Vilarino2007}, labeling noise was eliminated by recording gaze patterns only when an observer was looking at an ROI; observers had an On-Off switch to indicate when they were looking at an ROI. However, in practice, it is possible for users to forget to turn the switch on or off especially when burdened with high workloads. Furthermore, the need for manual input is bound to compromise on labeling speed.

None of the above studies have examined gaze based labeling of WSI data. We believe this is primarily because WSIs are not viewable using conventional image viewing software and substantial implementation effort is required to integrate WSI viewers with gaze tracking hardware. However, there are a limited number of studies that have investigated gaze patterns within the context of WSIs. 
In \cite{Brunye2017},  gaze patterns/eye movements were analysed to examine whether or not they can characterise the expertise level of pathologists. Another similar study was presented in \cite{Krupinski2006}. Both studies concluded that the gaze of expert/fully-trained pathologists converged much more quickly towards diagnostically relevant ROIs than junior or trainee pathologists. PathEdEx, an online tool that records and automatically groups the gaze patterns of pathologists was introduced in \cite{Ersoy2017}. The resulting clusters, along with clinical notes added by expert pathologists, were then employed for training of pathology students and trainees.
However, to the best of our knowledge, gaze patterns have so far not been utilized for labeling of WSI data for training deep neural networks. 

In addition to investigating the performance of gaze labelled data in a deep learning pipeline, we also examine the amount of time required for labeling of WSI data via gaze based annotations and compare it with the time taken using conventional handle labeling tools. Various gaze based and hand labeling annotation strategies are compared in \cite{Wang2017}. However, the discussion is primarily based on non-medical image applications. Given that WSIs contain a much larger amount of visual information (across multiple magnification levels) than natural images, it is imperative to thoroughly analyse the amount of time consumed for gaze based annotations of WSIs and compare it with the time required for conventional WSI annotation tools.

\section{Gaze Based Annotation}
\label{sec:GazeAnnotation}
In order to track a pathologist's gaze on screen we developed a purpose-built tracking software to enable the gaze tracking hardware to maintain a record of gaze patterns on WSIs, as the default tracking software provided by the hardware manufacturer was not compatible with WSI files.
Raw gaze patterns contained noise and some additional preprocessing steps were applied before using the gaze-labeling data for training an object detector. The different components of our gaze annotation pipeline are described in detail below.

\subsection{Experimental Setup}
\label{subSec:ExperimentalSetup}
The experiments conducted for this study utilized a hardware apparatus working in conjunction with a software for collection and analysis of data. The individual components of our experimental setup are described below.
\subsubsection{Hardware Setup}
\label{subsubSec:HardwareSetup}
For tracking of gaze patterns, we tested two different hardware.  The low cost and universal availability of webcams makes webcam based gaze tracking \cite{Papoutsaki2016} a very appealing option from a scalability perspective. Unfortunately, the performance of these was poor. This was observed across multiple webcams from different manufacturers. Web-cam based solutions were found to be widely inaccurate, were easily thrown off calibration by posture and lighting changes at the user end and hence not suitable for prolonged use in a single sitting. As reported in \cite{Jensen2019}, we found that infrared light based eye trackers present a variety of advantages such as more robust detection of the pupils.

\begin{figure*}
    \centering
     \subfigure[Locations of KPs/ROIs]{\label{fig:Raw2MapA}\includegraphics[trim={14mm 60mm 11mm 0},clip, height=45mm,keepaspectratio]{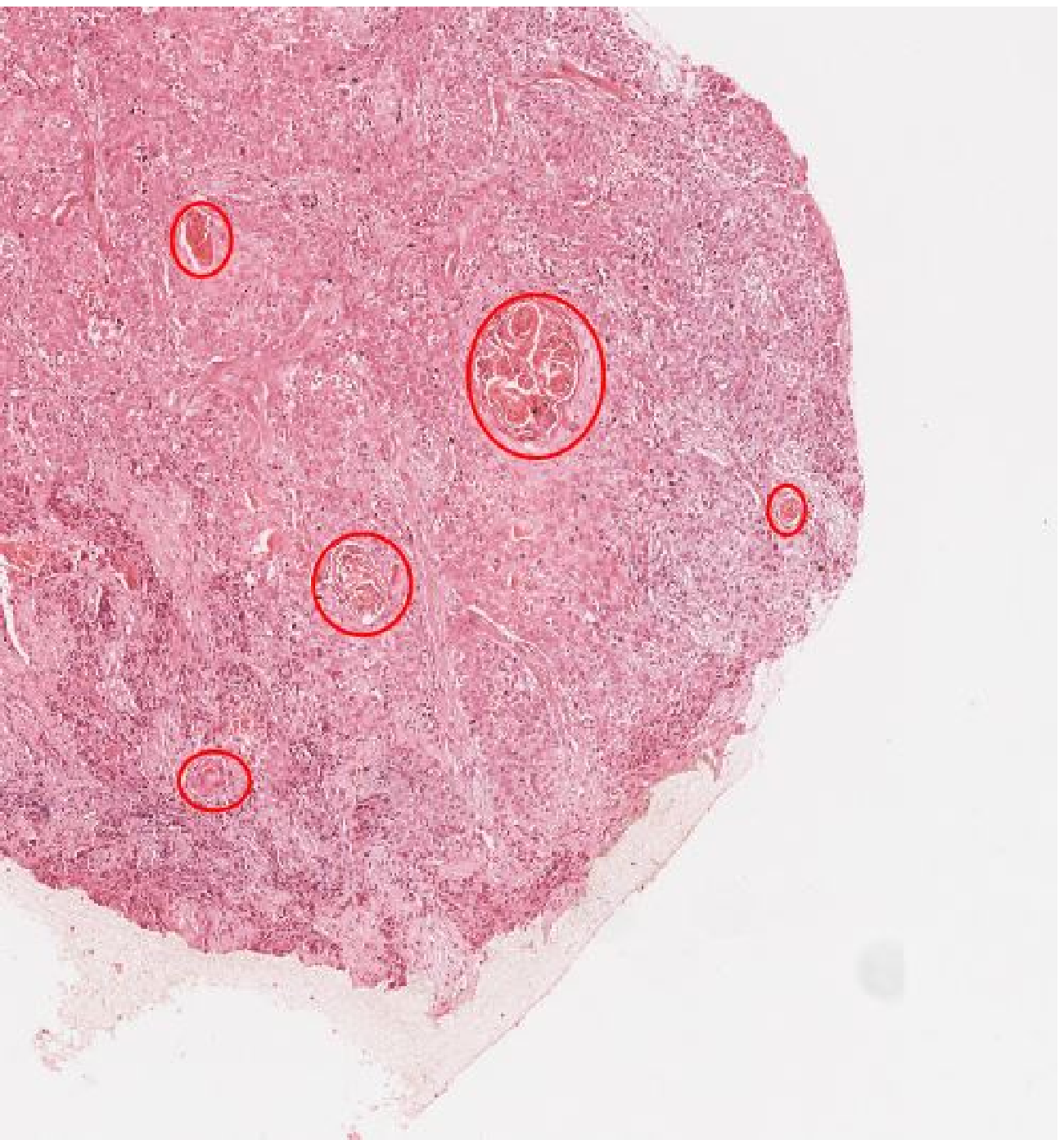}}
     \subfigure[Raw Gaze Points]{\label{fig:Raw2MapB}\includegraphics[trim={60mm 22mm 15mm 9mm},clip, height=45mm,keepaspectratio]{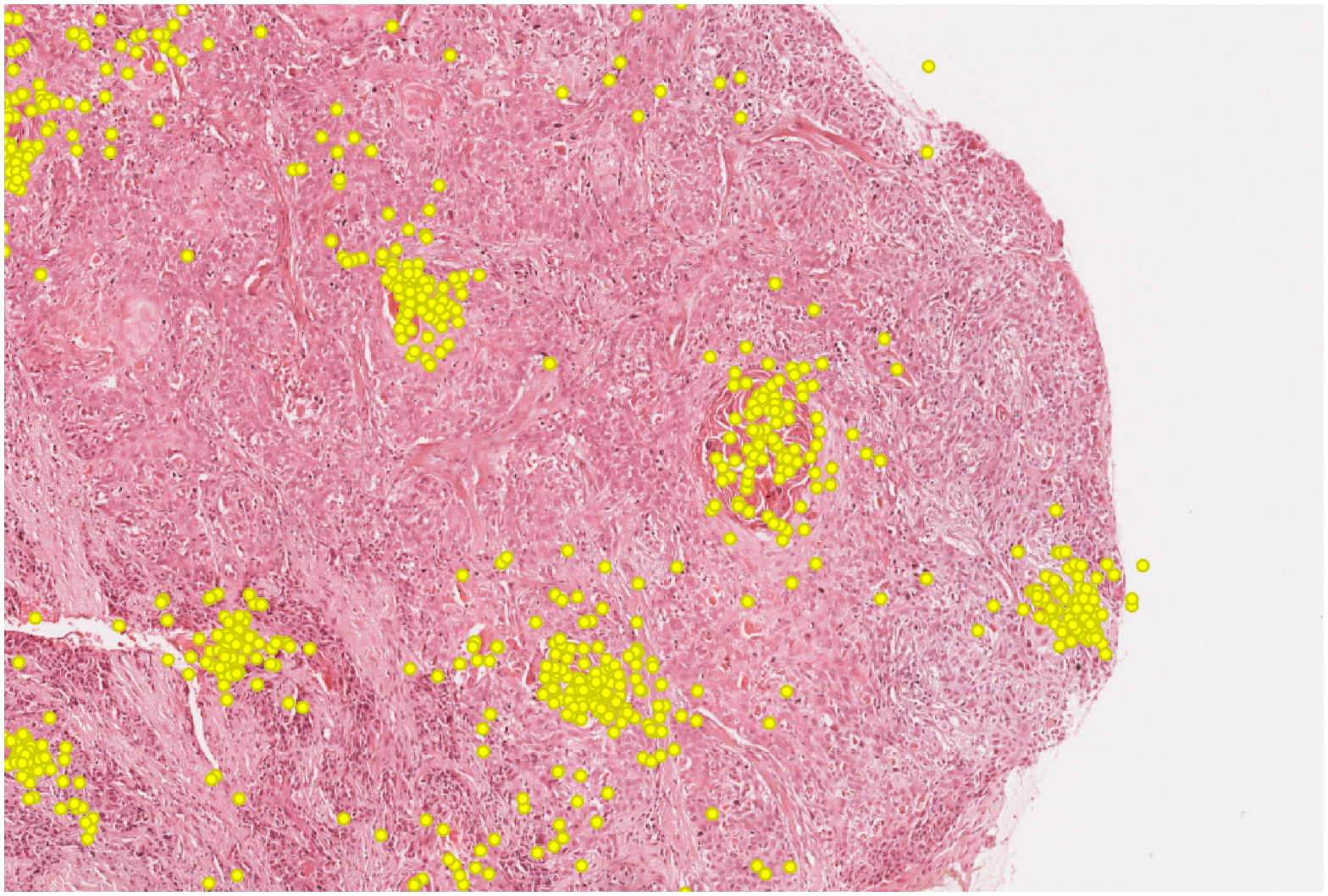}}
      \subfigure[Binary Map of ROIs obtained via KDE]{\label{fig:Raw2MapC}\includegraphics[trim={0 78mm 5mm 11mm},clip, height=45mm,keepaspectratio]{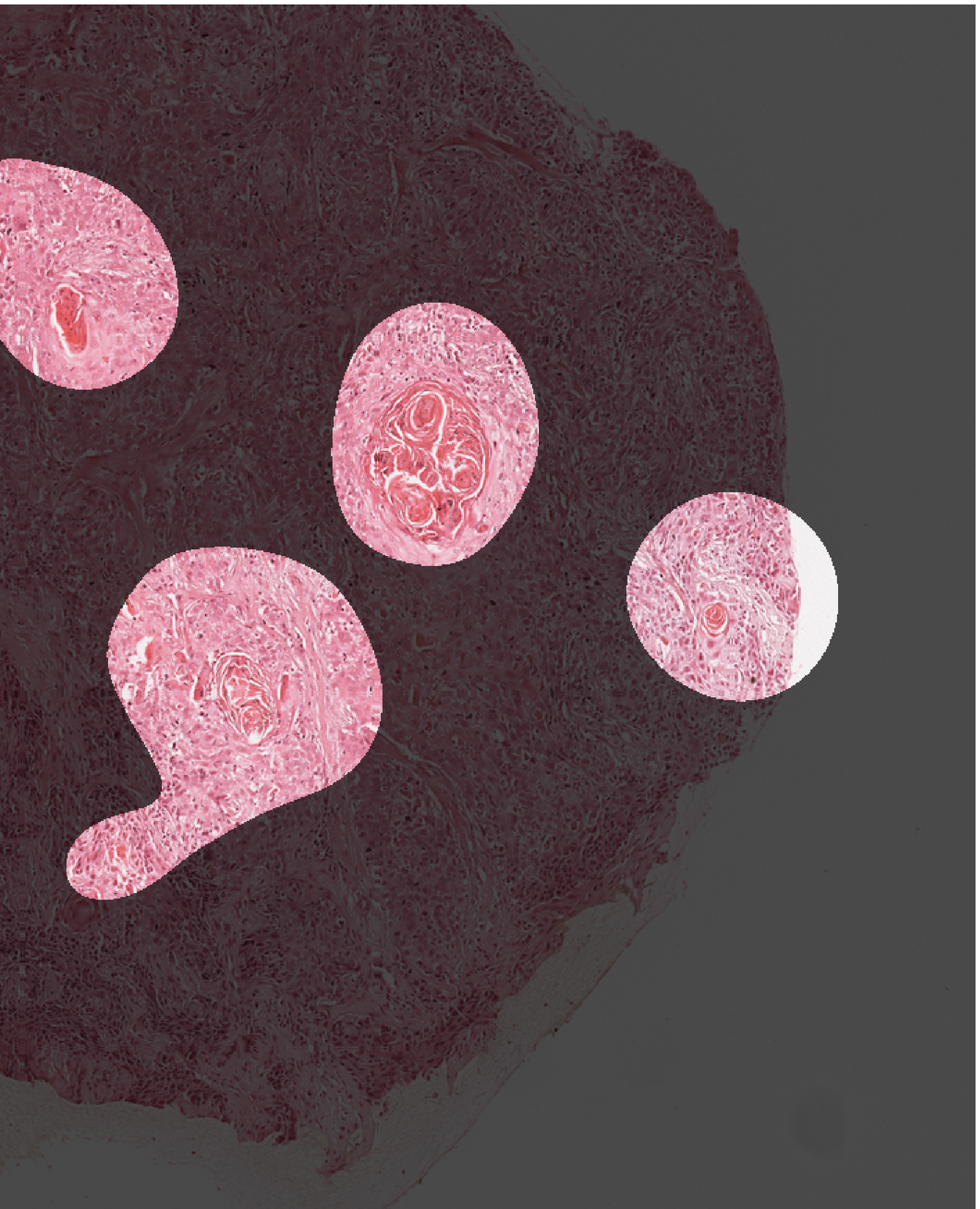}}
  \caption{Identification of ROIs on WSI using gaze patterns followed by KDE. Size of the KDE kernel can be changed to reduce the number of false patterns.}
  \label{fig:Raw2Map}
\end{figure*}

Our gaze detection hardware setup comprised a Gazepoint GP3 \cite{GazepointGP3}, a 26$''$ HD display monitor and a 1.6 GHz, Intel Core-i5, 8th Gen Desktop PC with 8GB RAM. The Gazepoint GP3 hardware is a desktop infrared gaze tracker with 60 Hz sampling frequency. The tracker was set up an arm's length away from the user's chair which is the distance recommended by the manufacturer for accurate calibration. 

\newcommand{\source}[1]{\caption*{Source: {#1}} }

\subsubsection{Software Setup}
\label{subsubSec:SoftwareSetup}
WSIs are stored in specialised formats which are not supported by conventional image viewing software. Annotation of WSI data typically requires access to either manufacturer specific software or open source tools such as  Qupath \cite{Bankhead2017}, Cytomine \cite{Maree2016} or Orbit \cite{Stritt2020}. However, all existing WSI viewers support only hand based annotations and currently do not have any gaze based annotation features. Consequently, we had to develop our own purpose-built software for gaze based annotations of WSI data. This required significant effort since we had to integrate the drivers of the gaze annotation hardware with Openslide which was used for reading the WSIs. Some of the main features of our gaze annotation tool include: WSI compatibility; Gaze recording; Visualization of gaze annotation; Noise mitigation via kernel density estimation; Generation of heatmaps and binary labels; Export of gaze annotation data to Qupath.
%\begin{itemize}
%  \item {WSI Viewer}
 % \item {Gaze Recorder}
  %\item {Gaze Annotation Visualiser}
  %\item {Noise Mitigation via Kernel Density Estimation \textcolor{red}{and Thresholding}}
 % \item {Generation of Heatmap and Binary Labels}
  %\item Scripts to Import Gaze Annotation Data to Qupath
%\end{itemize}

\subsubsection{Dataset Description}
\label{subSec:Dataset}
The dataset employed for the experiments in this work consisted of four Oral Squamous Cell Carcinoma (OSCC) Hematoxylin \& Eosin (H\&E) stained slides. These slides were digitized using an Aperio CS2 scanner. The generated WSIs are in RGB color space with a resolution of 0.4952$\mu m$ per pixel. Data collection for this study was approved by Riphah International University, Islamabad's Institutional Review board on Oct 10, 2019. Approval Number IIDC/IRC/2019/09/001. Data labeling and annotations were done by two expert pathologists. 
For training of object detection algorithms, we extracted 73 patches of resolution equal to $4000\times4000$ pixels with each patch containing at least one KP. Overall there were a total of 280 instances of KPs in our dataset. 85\% of the dataset (containing 231 instances of KPs) was randomly selected for training/validation. The remaning 15\% (containing 52 instances of KPs) was set aside for use as a hidden test set.

\subsection{Noise Mitigation via Kernel Density Estimation} 
\label{subSec:KDE}

In our experiments, two pathologists were asked to search for ROIs (KPs) on the WSI displayed on the screen in front of them. They were asked to focus on each ROI they located for 1 to 2 seconds. ROIs were identified using gaze intensity on the WSI. However, identification using raw data resulted in false positives since there were multiple instances of fixation on regions other than ROIs as well. Therefore, kernel density estimation (KDE) was employed to cluster (and threshold) gaze points belonging to true ROIs and differentiate them from noise \cite{Parzen1962}. An example of this annotation process is provided in Fig.~\ref{fig:Raw2Map}. The baseline hand annotated WSI with ROIs indicated by red ovals is shown in Fig.~\ref{fig:Raw2MapA}. The corresponding raw gaze points recorded by the gaze tracking hardware are shown in Fig.~\ref{fig:Raw2MapB}; this raw data contains multiple gaze points that lie outside of the true ROIs. The binary mask obtained after application of the KDE is shown in Fig.~\ref{fig:Raw2MapC}, it can be observed that KDE removes noisy gaze points. 

\begin{figure}
\centering
\includegraphics[width=0.5\textwidth]{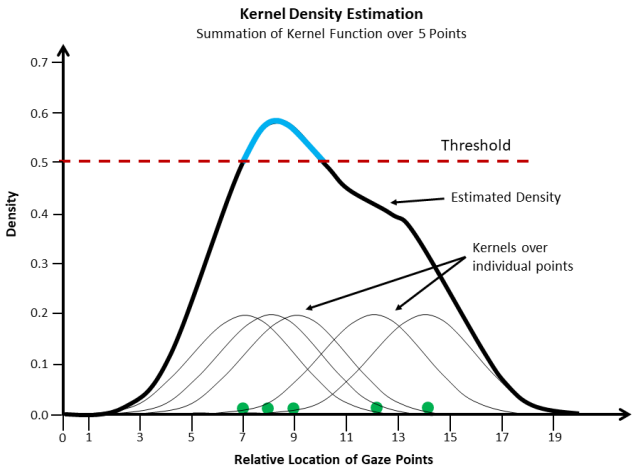}
  \caption{2D view of kernel density estimate of 5 gazepoints with binary colour map obtained after 0,1 thresholding. Regions above the threshold are considered part of an ROI. Threshold value and kernel size enable us to vary the size of ROIs.}
  \label{fig:kde}
\end{figure}

KDE was deemed suitable for our target ROIs since KPs tend to have, more or less, circular or elliptical shapes and thus can be easily captured by fitting 2D Gaussian kernels to the raw gaze points. When two or more Gaussian kernels overlapped, their values were added together resulting in an enhanced fixation intensity. In areas with a high density of gaze points the magnitude of fixation intensity was higher. The higher the magnitude, the more distinguished a fixation region was in the resulting binary mask representation. The KDE algorithm takes in to account two parameters: \textit{size} ($\sigma$) and \textit{threshold} ($\tau$). The size (or standard deviation) of the Gaussian kernel controls the expanse of the neighbourhood of points that form a cluster. The threshold ($\tau$)
was employed to discard small sized clusters; any cluster (or part of a cluster) with gaze density below $\tau$ was discarded and not included in the final label mask. A simple illustration of KDE and the role of thresholding is provided in Fig.~\ref{fig:kde}.

\begin{figure}[t]
\centering
\includegraphics[width=\columnwidth]{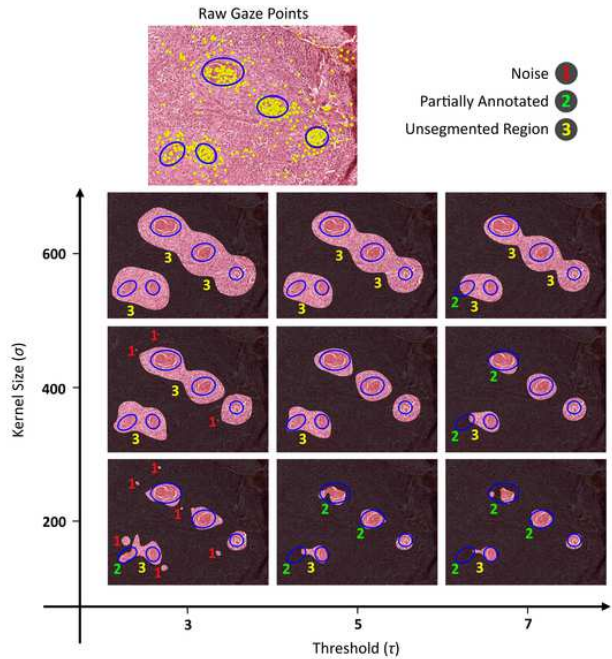}
\caption{Impact of Thresholding and Kernel Size: Higher threshold values reduce outliers or noisy data points. Large kernel sizes merge closely spaced, small ROIs. Small kernel sizes can break large ROIs into multiple smaller ROIs.}
 \label{fig:KSvsTHE}
\end{figure}

Parameters of the KDE algorithm ($\tau$ and $\sigma$) can be employed to control the number of outliers and size of the ROIs. More specifically, assigning a larger value to $\tau$ eliminates outliers/noisy gaze points.
The impact of threshold $\tau$ and kernel size $\sigma$ is illustrated in Fig.~\ref{fig:KSvsTHE}. Use of larger thresholds eliminates outliers/noisy gaze points but may also result in missing small sized ROIs. The kernel size mainly has an impact on the size of the ROI detected, larger values of $\sigma$ result in detection of large sized ROIs whereas smaller values are preferable when the target ROIs are small sized. It can be observed in Fig.~\ref{fig:KSvsTHE} that larger values of $\sigma$ are suitable for large ROIs but can cause multiple, closely spaced, small ROIs to merge together into a single ROI. By contrast, smaller $\sigma$ values are suitable for small ROIs but can split a single, large ROI into multiple smaller ROIs. %This means that in cases with large size variations, a single value of $sigma$ may not work since it may be either, too large for small ROIs or, too small for large ROIs. 
We mitigated this problem by employing smaller values of $\sigma$ for cases containing smaller ROIs and switching to larger values for cases containing larger ROIs. For cases containing ROIs of different sizes, we employed two different values of $\sigma$ on gaze data (from a single recording session) and then merged the two binary masks. 
More details about the impact of different $\sigma$ and $\tau$ values on the accuracy of gaze-labels are provided in section~\ref{sec:Results} where we present the mean Intersection-Over-Union (mIOU) between the (ground-truth) Hand generated masks and the (KDE-processed) Gaze generated masks as a function of $\sigma$ and $\tau$.

\subsubsection{Comparison of Hand and Gaze Generated Maps}
\label{subsubSec:mIOUhandGaze}

To evaluate the quality of the gaze based maps (generated by KDE), we compared them to the hand labelled maps. More specifically, we evaluated the mIOU between gaze labelled maps and their corresponding (ground-truth) hand labelled maps for various values of $\sigma$ and $\tau$. The IOU is a performance evaluation metric that is widely employed in semantic segmentation tasks; it is defined in the context of our application as,
\begin{equation}
    IOU = \frac{M_H \cap M_G} {M_H \cup M_G}
\end{equation}
where $M_H$ and $M_G$ denote the (ground-truth) hand  and the gaze labelled masks, respectively.
The mean and standard-deviation of the IOU between $M_H$ and $M_G$ masks were used to examine the impact of the KDE parameters ($\sigma$ and $\tau$) on the quality of the gaze labelled maps and identify the parameter values that result in gaze labelled masks that have the highest overlap with the hand labelled masks. 
Due to variations in sizes of ROIs in different images, the threshold $\tau$ was adjusted based on the statistics of data within each image. Selection of the threshold parameter for each image is described below.

Given a kernel size $\sigma$, KDE was applied to raw gaze data of the slide image. Application of KDE resulted in a set of $c$ clusters represented by 
\begin{equation}
\label{eq:clusterPDFs}
    f^i_{X,Y}(x,y) \qquad \text{where,}~~i\in [1,\dots, c]
\end{equation}
Here, $f^i_{X,Y}(x,y)$ denotes the probability density function of the $i$-th cluster. The number of clusters, $c$, is automatically selected by KDE and depends primarily on the kernel size ($\sigma$). As stated previously, smaller values of $\sigma$ result in a larger number of (small-sized) clusters and vice versa. We then evaluated the sample mean of each individual pdf in (\ref{eq:clusterPDFs}),
\begin{equation}
\label{eq:SampleMean}
    \bar{\theta}_i = \frac{1}{b_i}\sum_{j=1}^{b_i}f^{i,j}_{X,Y}(x,y)
\end{equation}
where $b_i$ denotes the number of bins in $f^i_{X,Y}(x,y)$. It is to be noted here that $\bar{\theta}_i$ is the sample mean evaluated over the magnitude ($z$-axis) of the cluster's pdf; it should not be confused with the cluster's mean vector, $\boldsymbol{\mu}^i_{X,Y}$. Furthermore, $f^{i,j}_{X,Y}(x,y)$ denotes the value of the $j$-th bin of the pdf of the $i$-th cluster. In order to apply a single threshold across the entire image, we again took the sample mean as below,
\begin{equation}
\label{eq:MeanofMean}
    \bar{\theta} = \frac{1}{c}\sum_{i=1}^{c}\bar{\theta}_i \quad \cdot
\end{equation}
In order to be able to vary the threshold we employed a normalized and scaled version of $\bar{\theta}$ as below,
\begin{equation}
    \label{eq:NormThresh}
    \tau = n\times \frac{\bar{\theta}}{m} 
\end{equation}
where $n$ denotes the scaling factor and $m$ denotes the maximum value observed across all bins of all cluster pdfs
\begin{equation}
    \label{eq:maxVal}
    m = max\left( f^{i,j}_{X,Y}(x,y)\right) \qquad\qquad \forall~~(i, j)\cdot
\end{equation}
Detailed results on the mIOUs between $M_H$ and $M_G$ for different values of the kernel size $\sigma$ and scaling factor $n$ are presented in section~\ref{sec:Results}. The highest mIOU value that we observed was 0.6714. In this paper, we focus on the task of object detection. Good performance of object detection would indicate that the KDE gaze masks contain enough information to enable estimation of ROI size and location within a slide image.

\subsection{KP Detection}
\label{subSec:Objdet}
We used the binary masks obtained after application of KDE to train object detectors for localization of KPs in WSI data. The binary masks output from the KDE block are oval/circular in shape whereas object detectors generally require square/rectangular shaped objects. Therefore, the irregular shaped masks were converted into tightly fitting bounding boxes before being input to the object detector which was then trained with the input label masks using transfer learning. Three popular object detection algorithms were employed in our experiments: (1) Faster R-CNN  with Inceptionv2 architecture (2) YOLOv3 with DarkNet53 architecture (3) YOLOv5 with DarkNet53 architecture.

%\begin{figure}
%    \centering
%     \subfigure[Threshold Value 0.5]{\label{fig:a}\includegraphics[ height=35mm,keepaspectratio]{threshold-0.5.jpg}}
%     \subfigure[Threshold Value 0.75]{\label{fig:b}\includegraphics[trim={0 0 0 3mm},clip, height=35mm,keepaspectratio]{threshold-0.75.jpg}}
%  \caption{Impact of Thresholding: Higher threshold values reduce outliers or noisy data points.}
%  \label{fig:Threshold}
%\end{figure}

\begin{table*}[tbp]
	\renewcommand{\arraystretch}{1.5} %Increase Spacing b/w rows (default is 1)
	\setlength{\tabcolsep}{9pt} %Increase col width (default is 6pt)
	\caption{Time Taken by Pathologist A \& B for Each Annotation Method}
	\centering
	\begin{tabular}{|c|c|c|c|c|c|c|c|c|c|}
	\hline
	{} & \multicolumn{3}{c}{\textbf{Freehand}} & \multicolumn{3}{|c}{\textbf{Bounding Box}} & \multicolumn{3}{|c|}{\textbf{Gaze}}\\
	\hline
	{\textbf{Pathologist}} & \textbf{A} & \textbf{B} & \textbf{Avg. Time} & \textbf{A} & \textbf{B}&  \textbf{Avg. Time} & \textbf{A}& \textbf{B} & \textbf{Avg. Time}\\
	\hline
    	\textbf{Total Time (mm:ss)} & 11:32 & 15:24  & 13:28 & 04:50 & 04:41 & 04:46 & 01:49 & 02:13 & \textbf{02:01}\\
	\hline
	\textbf{Total Time in Seconds} & 692 & 924 & 808 & 290 & 281 & 285.5 & 109 & 133 & \textbf{121}\\
	\hline
	\textbf{Average Time/KP in Seconds} & 13.84 & 18.48 & 16.16 & 5.80 & 5.62 & 5.71 & 2.18 & 2.66 & \textbf{2.42}\\
	\hline
	\end{tabular} 	
	\label{tab:Time Taken}
\end{table*}

During training we trained two sets of the above listed object detectors. Set-1, denoted by $S_G$, trained using the Gaze generated label masks, $M_G$. Set-2, denoted by $S_H$, trained using the (ground-truth) hand labelled masks, $M_H$. $S_H$ was trained for benchmarking purposes. During testing, the prediction of each object detector was compared with the ground-truth masks, $M_H$, regardless of whether the training labels came from $M_G$ or $M_H$. The training protocols, architecture, hyperparameters and configurations such as patch size, number of iterations, learning rate, batch size, input image resolution were identical for both object detector sets, $S_G$ and $S_H$. The only difference was in the labeling strategy employed for annotating the training data. %Both detectors were fine-tuned using an initial learning rate of $2 \times 10^{-4}$ and batch size of 2. At the end of training, Faster R-CNN's multi-task loss function plateaued at a value of 0.06672 for the gaze annotated dataset and 0.03229 for the hand annotated dataset.
Both detector sets were evaluated on the same test set using the following metrics: Precision, Recall (Sensitivity), Mean Average Precision (mAP), Miss-Rate, False-Positives-Per-Image (FPPI) and log-Average Miss Rate (LAMR). For thorough and unbiased assessment, all metrics were evaluated at different values of overlap between detected and ground truth bounding boxes. This overlap is typically quantified via the intersection over union between the detected and ground truth bounding boxes $(IOU_{dt})$ \cite{dollar2011pedestrian} which is defined as:
\begin{equation}
    IOU_{dt} = \frac{area\left(BB_{dt}\cap BB_{gt}\right)}{{area\left(BB_{dt}\cup BB_{gt}\right)}}
    \label{eq:IOU}
\end{equation}
where $BB_{dt}$ and $BB_{gt}$ denote the detected bounding box and the ground truth bounding box respectively. A detection is considered successful if the $IOU_{dt}$ exceeds a predefined overlap threshold $(OT)$. In our experiments, $OT$ was varied between 0.1 to 0.95. 
%\begin{eqnarray}
%\label{eqn:Metrics1}
%precision & = &\frac{TP}{TP + FP} \\
%& ~ & \nonumber \\
%recall & = &\frac{TP}{TP + FN}% \\
%\end{eqnarray}
%
%\begin{eqnarray}
%\label{eqn:Metrics2}
%F1\text{-score}  & = & 2\cdot \frac{precision \cdot recall}{precision + recall} 
%\end{eqnarray}
%where $TP$ denotes the number of `True Positive' predictions by the object detector, $FP$ denotes the number of `False Positive' predictions and $FN$ denotes the number of `False Negative' predictions.

WSIs have very high resolution and large size, making it difficult for them to be processed as a whole by a neural network. Therefore, we divided each WSI into a set of uniformly sized  $4000 \times 4000$ patches before feeding them to a neural network for training/testing. The resolution of each patch was left unchanged to minimise loss of contextual information and retain morphological features. The patch size was selected so that the patches remained large enough to subsume large KPs and small enough to be digestible by the neural network.

%\begin{figure}
%    \centering
%     \subfigure[Kernel Size 1600]{\label{fig:a}\includegraphics[trim={0 0 30mm 0},clip, height=35mm,keepaspectratio]{Kernal Size 1600.jpg}}
%     \subfigure[Kernel Size 400]{\label{fig:b}\includegraphics[height=35mm,keepaspectratio]{Kernel Size 400.jpg}}
%  \caption{Impact of Kernel Size: Large kernel sizes merge closely spaced, small ROIs. Small kernel sizes can break large ROIs into multiple smaller ROIs.}
%  \label{fig:KernelSize}
%\end{figure}

\subsection{Comparison of labeling Time}
\label{subSec:LabelTime}
The primary motivation for replacing hand annotations with gaze based annotations is potential saving in time and effort required for data labeling. Although it does seem obvious that gaze based annotations save time, the vast majority of the existing studies on gaze based annotations of biomedical data have not quantified the amount of time that is saved in comparison to hand annotations. This information should be available in order to make future improvements in gaze based data annotation systems like ours. Some data is available for natural images; for example, according to \cite{Su2012}, the median time for drawing a bounding-box during a crowd-sourcing effort to label ImageNet dataset was observed to be 26 seconds. For segmentation the average labeling time for a single image was 15-60 minutes for the MSRC dataset \cite{Papadopoulos2014}. For gaze labeling, the average time for one object per image (in the 2012 Pascal VOC challenge dataset) was around 1 second \cite{Kohli2008}. Given the inherent differences between natural and medical images, we believe it is important to quantify annotation times for hand and gaze based labeling for medical images. Therefore,  we conducted a set of dedicated experiments in order to quantify the amount of time consumed by each data labeling strategy. For gaze annotations, we calculated the average amount of time consumed for labeling the shape of a single ROI. For hand annotations, we calculated the average time consumed for labeling a single ROI using two distinct approaches: (1) \textit{Bounding-Box} labeling, where the objective was to put a bounding box around a target ROI (2) \textit{Freehand} labeling, where the objective was to draw a contour around an ROI. Both the preceding hand annotation strategies were performed on a desktop computer with a mouse using the QuPath software. The pathologists did have prior experience of working with QuPath and were familiar with its user interface.

Among these three strategies, freehand annotation consumes the most time but also provides the most information since it captures the exact shape and size of ROIs whereas bounding-box and gaze annotations are quicker but provide a comparatively lower amount of information since they capture only the approximate shape and size of ROIs. The bounding  box  annotation includes the time taken for drawing  the  bounding  box  itself  and  re-adjusting  the  anchor points for a tightly fitting box around the ROI. The default settings of Qupath were used for our experiments that require the user to re-select the annotation tool after each ROI annotation. Hand annotation measures both exact and approximate shape, whereas our current gaze labeling strategy measures only approximate shape and size.
For  gaze  annotation,  the  annotators  were  instructed  to  fixate or  hold  their  gaze  on  a  Keratin  Pearl  for  what  they  felt  was at least 2 seconds.  Annotators were allowed to practice holding their gaze for a few seconds on our software on a few samples before commencing gaze annotations.

\section{Results and Discussion}
\label{sec:Results}
The results of all the experiments described in section~\ref{sec:GazeAnnotation} are presented in chronological order below. This is followed by a discussion of the salient results.
\subsubsection{Hand and Gaze Map Comparison}
\label{subsec:IOUcomparison}
\begin{figure}[t]
\centering
\includegraphics[width=\columnwidth]{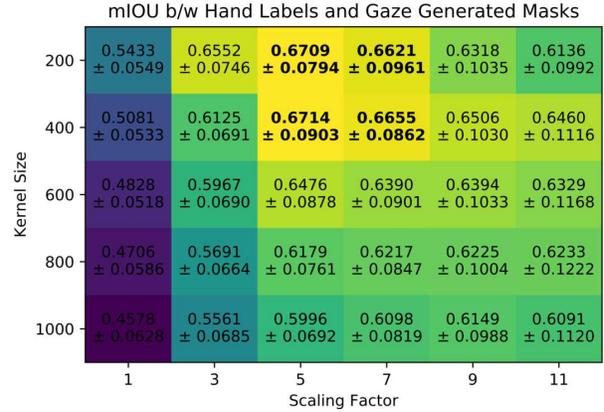}
  \caption{mIOU as a function of Kernel Size and Scaling Factor}
  \label{fig:mIOU_img}
\end{figure}

The mIOUs between $M_H$ and $M_G$ for different values of the kernel size, $\sigma$, and scaling factor, $n$, are shown in Fig.~\ref{fig:mIOU_img}. The mIOU is plotted as a function of the scaling factor $n$ instead of the normalized threshold $\tau$ because $\tau$ depends on the statistics of the ROIs within an image and thus can be different from one image to another, whereas $n$ is same across all images (for one setting) and therefore, more suitable for comparison purposes.
Fig.~\ref{fig:mIOU_img} also shows the standard deviation of the IOU observed for each parameter setting. The highest value of $\text{mIOU} = 0.6714$ is observed at $\sigma=400$ and $n=5$. In terms of kernel size, it seems that smaller values ($\sigma=200$ and $\sigma=400$) result in better mIOUs than larger values. This makes sense, since larger values of $\sigma$ tend to merge neighbouring clusters. In terms of the scaling factor $n$, it seems that the highest mIOU values are observed around values of $n=5$ and $n=7$. It is also noted that the standard deviation of mIOU is higher at larger values of $n$. This is most likely due to smaller clusters (or parts of clusters) being missed as $n$ (or the threshold $\tau$) is increased. The analysis presented in Fig.~\ref{fig:mIOU_img} indicates that $\sigma=400$ and $n=5$ result in the best gaze masks.  
%In this study, we chose to examine the performance of gaze labeling in an object detection setting to examine whether gaze masks contained enough information to enable estimation of ROI size and location within a slide image. 

\subsubsection{KP Detection}
\label{subsubSec:KPresults}
\begin{figure*}[t]
\centering
\centering
  \subfigure{\includegraphics[width=.4\textwidth]{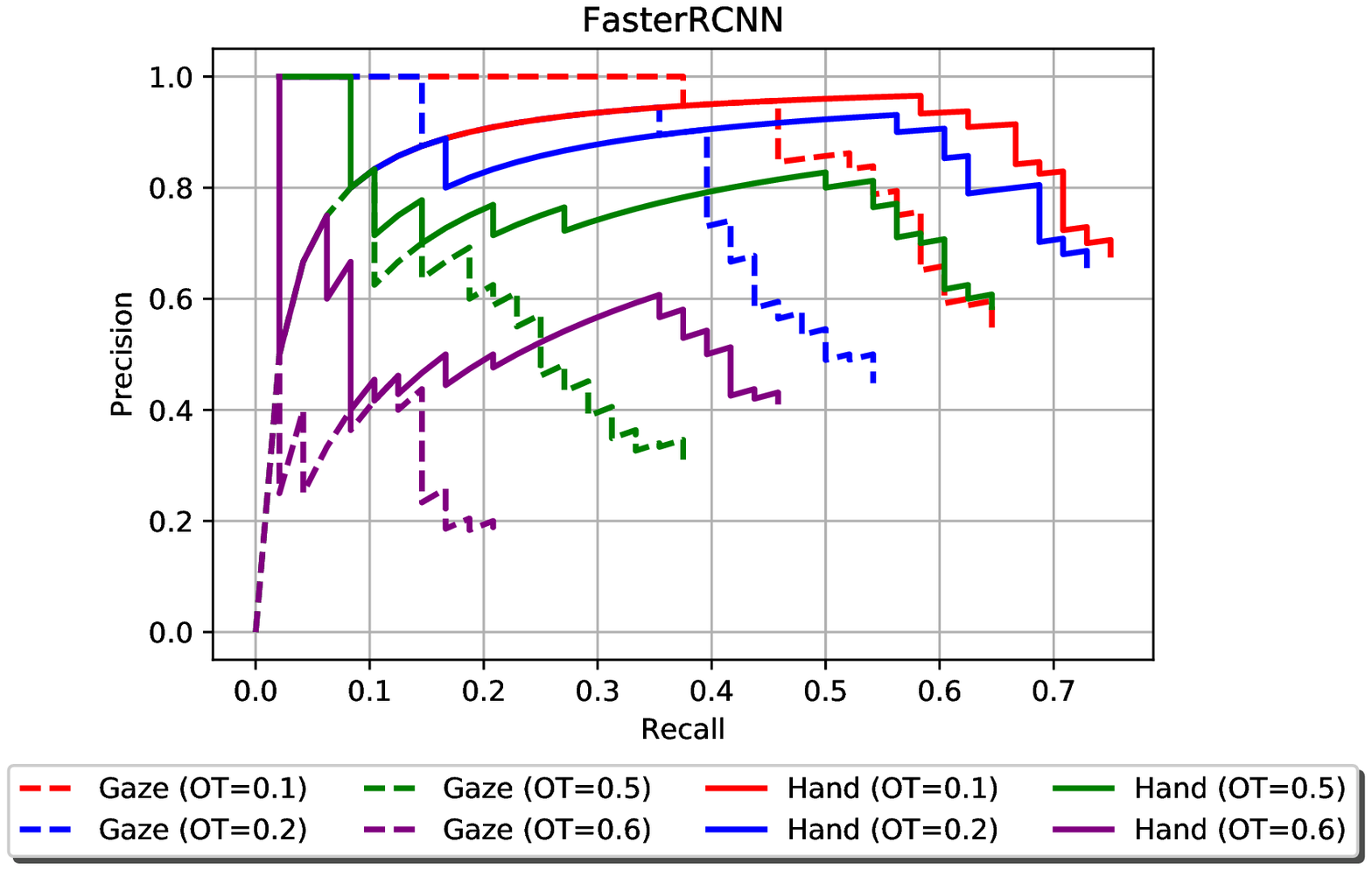}}\quad
  \subfigure{\includegraphics[width=.4\textwidth]{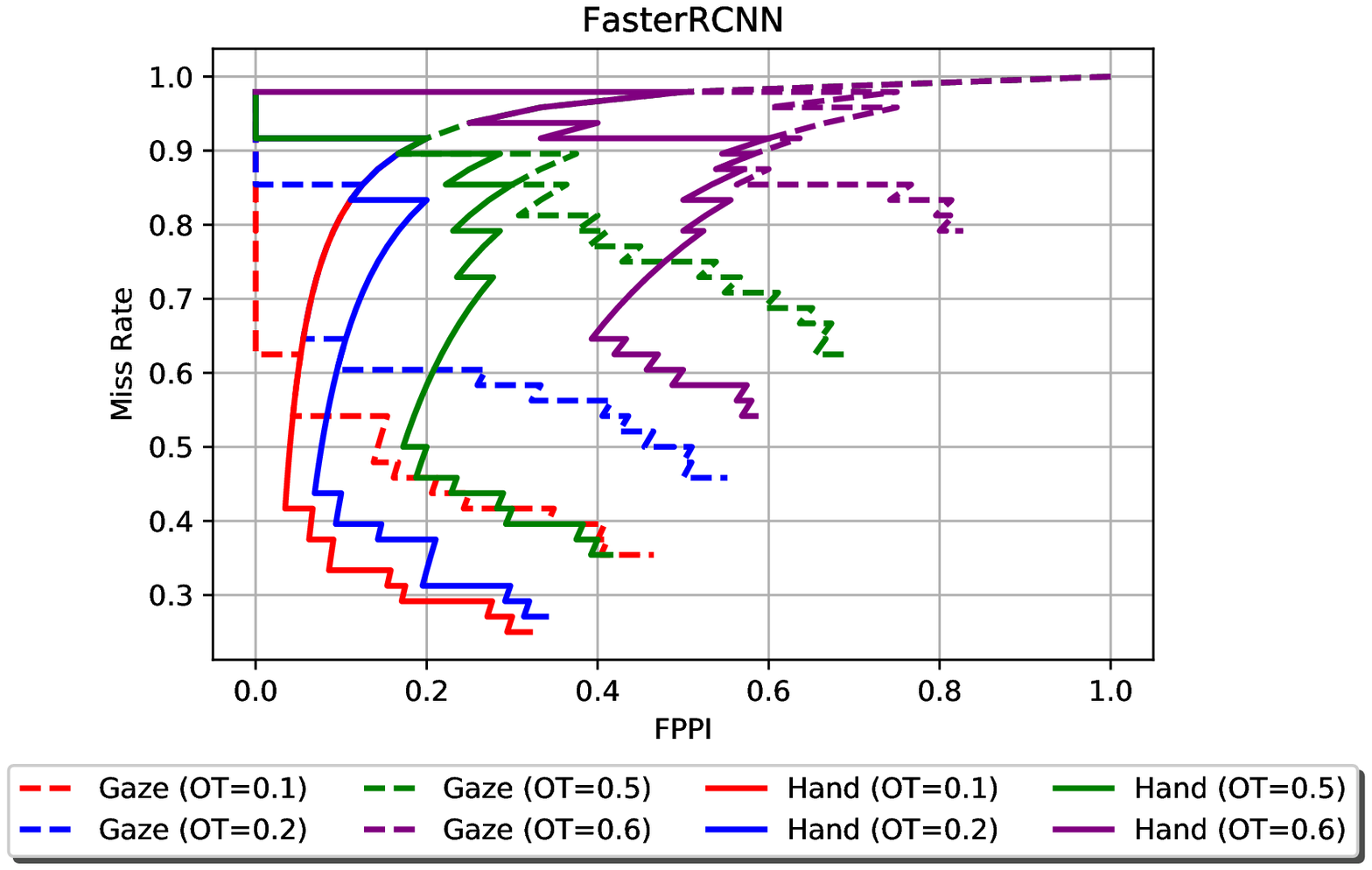}}\\
  \subfigure{\includegraphics[width=.4\textwidth]{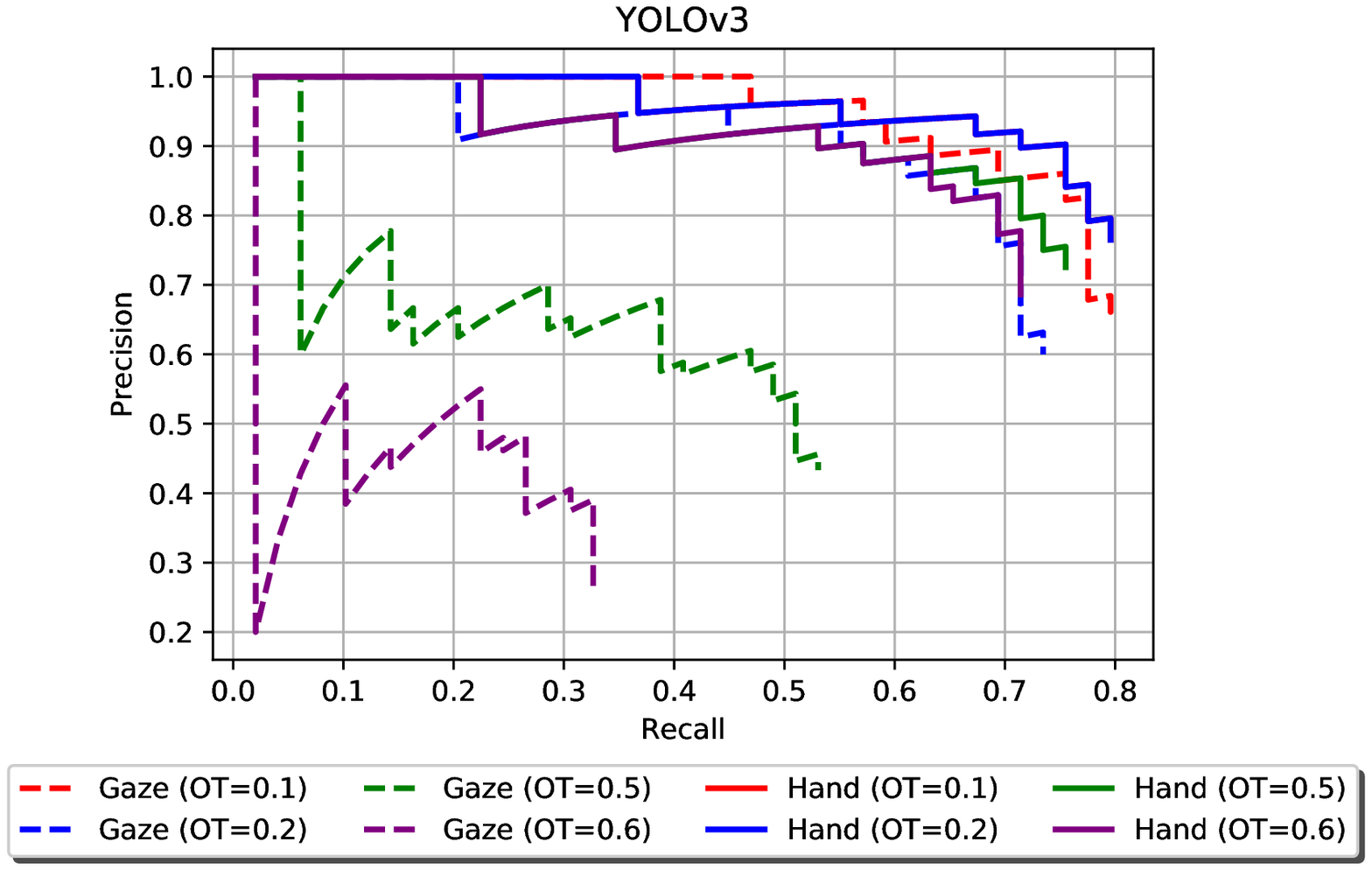}}\quad
  \subfigure{\includegraphics[width=.4\textwidth]{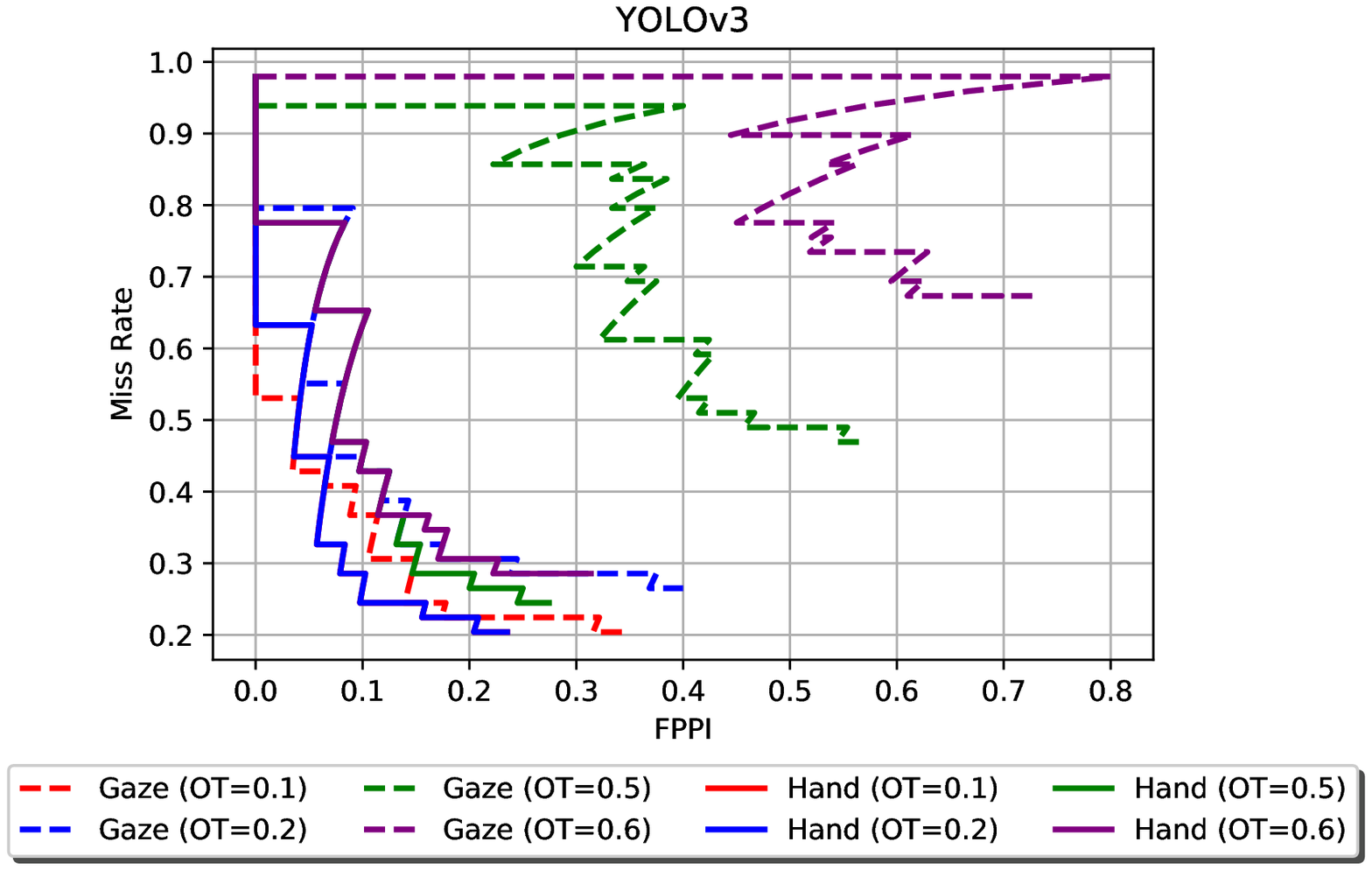}}\\
  \subfigure{\includegraphics[width=.4\textwidth]{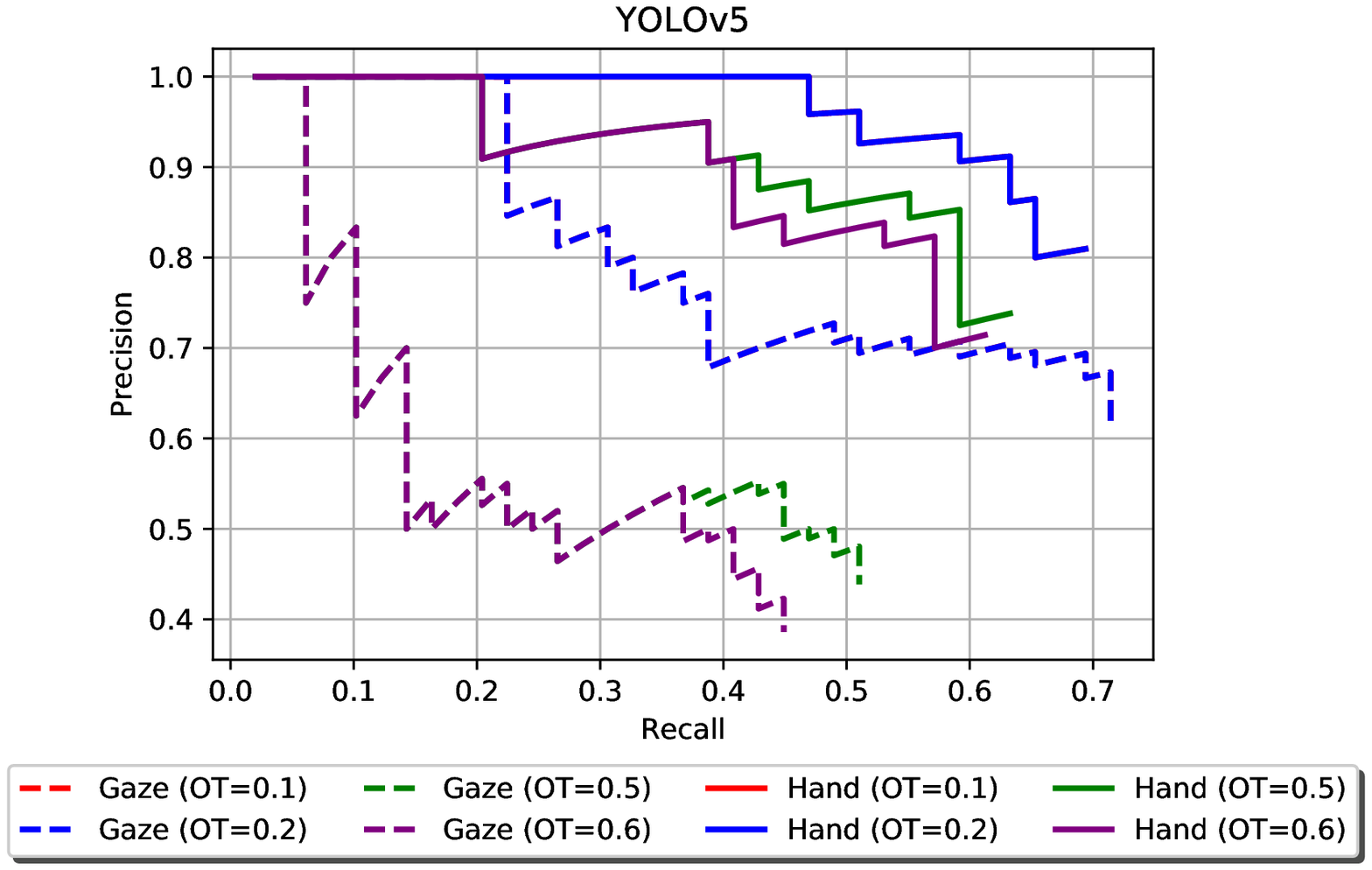}}\quad
  \subfigure{\includegraphics[width=.4\textwidth]{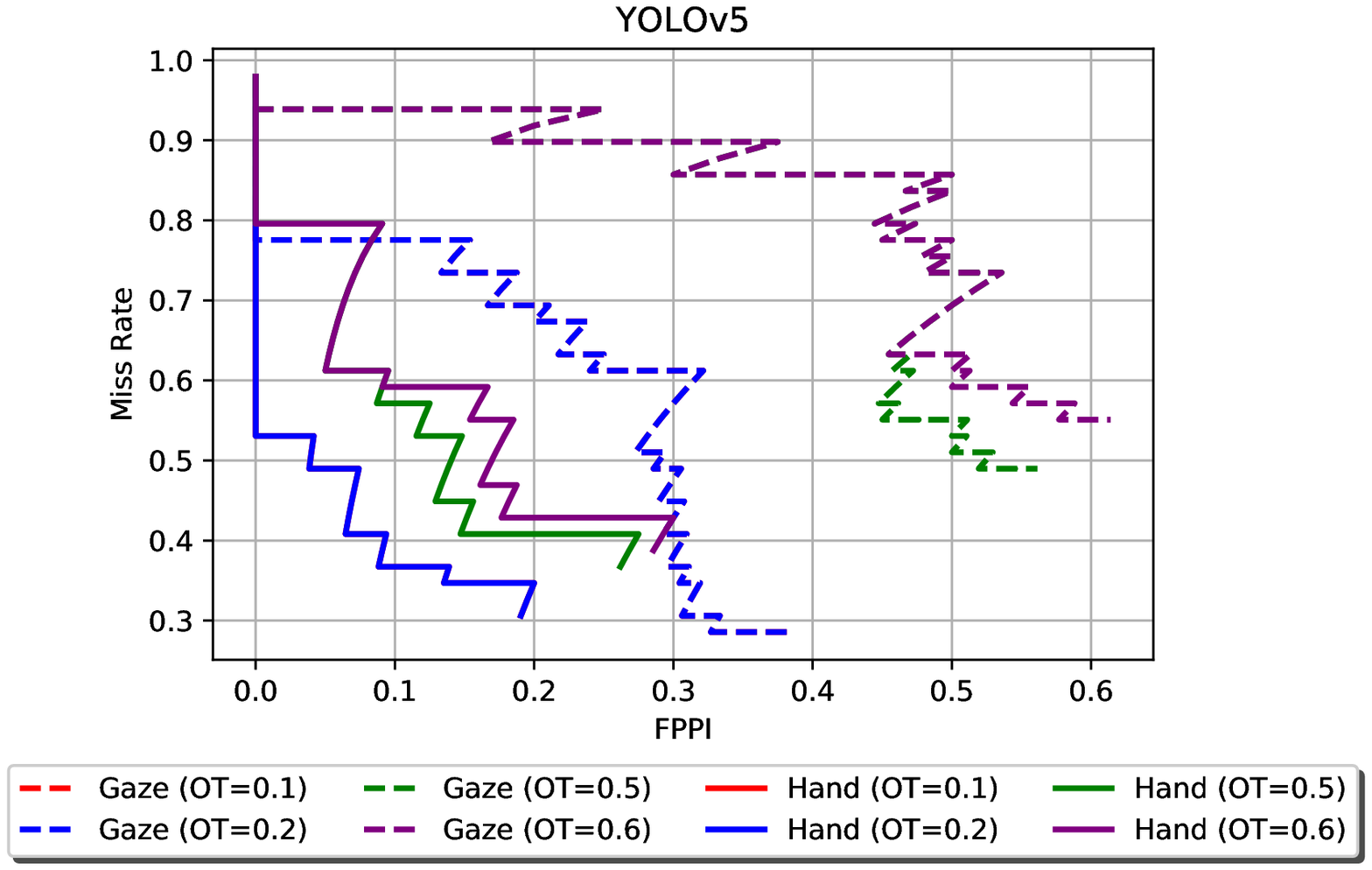}}
\caption{Precision versus Recall curves and Miss-Rate versus False-Positive-Per-Image (FPPI) curves, at different Overlap Threshold (OT) values, for three different object detectors. Dashed lines correspond to object detectors trained using Gaze labelled data. In some cases the observed curves are identical and thus appear as one since they are exactly aligned. For example, for the YOLOv3 hand detector the Precision-Recall curve for $OT=0.1$ is not visible because the curve for $OT=0.2$ lies exactly on top of it.}
 \label{fig:PrecRec_MRFPPI}
\end{figure*}

The Precision versus Recall and the Miss-Rate versus FPPI curves observed on the trained object detectors for different values of $OT$ are presented in Fig.~\ref{fig:PrecRec_MRFPPI}. 
The set of benchmark object detectors, $S_H$,  provide an estimate of the performance that can be obtained on this problem via training using the conventional hand labelled approach. Performance of all  detectors may be improved further by enhancing the size of the dataset and/or employing better detection approaches. Another factor limiting performance may be `Border effects' that arise when  WSI images are split into smaller image to make deep learning processing possible. %Such effects could possibly be the underlying reason for the FN visible in the bottom row of Fig.~\ref{fig:Results}. 
Impact of border effects may be mitigated by having overlap between the tiles with an amount roughly corresponding to the diameter of a typical ROI and then merging detection from the different tiles covering the same region on the WSI. However, overlap based approaches will require extra processing time, the amount of which will depend on the diameter of the largest expected ROI. We have left these measures for future work since maximizing the detection performance is not the primary objective of this work. The plots in Fig.~\ref{fig:PrecRec_MRFPPI} demonstrate that the performance of all detectors degrades with increasing values of $OT$. In both $S_H$ and $S_G$ the YOLOv3 is the best detector. For all gaze trained detectors, in $S_G$, performance degrades substantially at high values of $OT$. However, at low values of $OT$, YOLOv3 and YOLOv5 in $S_G$ deliver performance similar to those observed by their corresponding detectors in $S_H$ . 
%It is highlighted that in some cases the observed curves are identical and thus appear as one since they are exactly aligned. For example, for the YOLOv3 hand detector the Precision-Recall curve for $OT=0.1$ is not visible because the curve for $OT=0.2$ lies exactly on top of it. }
%In, $S_H$ the best performance is achieved by the YOLOv4 architecture $(F1\text{-score} = 0.7447)$, followed by the Faster R-CNN $(F1\text{-score} = 0.7349)$ and the YOLOv3 architecture $(F1\text{-score} = 0.6772)$. 
%For the set of gaze trained object detectors, $S_G$, the best performance is achieved by Faster R-CNN $(F1\text{-score} = 0.7126)$, followed by YOLOv4 $(F1\text{-score} = 0.7059)$ and YOLOv3 $(F1\text{-score} = 0.6250)$. 
%This means that, in terms of $F1\text{-score}$ the gap between the best performing detector in $S_H$ and the best performing detector in $S_G$ is equal $0.0321$. Whereas, the gap between the worst performing detectors in $S_H$ and $S_G$ is equal to $0.0522$. 
%It is highlighted that corresponding detectors in $S_H$ and $S_G$ were tested on the exact same WSI test set. 

\begin{figure}[t]
\centering
\includegraphics[width=\columnwidth]{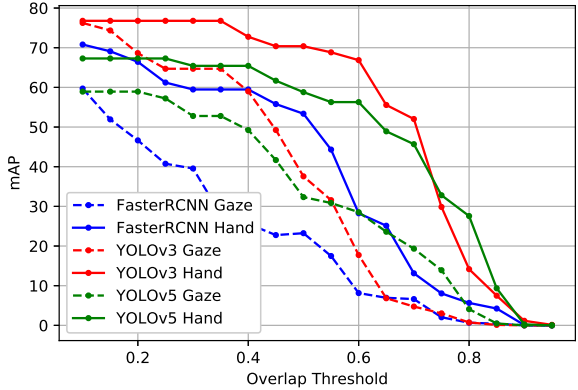}
\caption{Mean average precision (mAP) of all detectors observed at different overlap threshold settings.}
 \label{fig:mAPvsOT}
\end{figure}

\begin{figure}[t]
\centering
\includegraphics[width=\columnwidth]{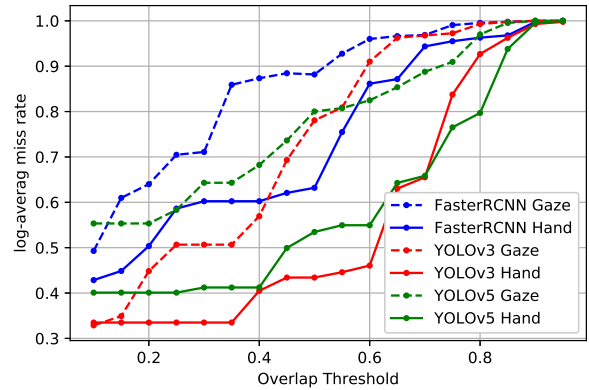}
\caption{log-average miss rate (LAMR) of different detectors observed at different overlap threshold settings.}
 \label{fig:LAMRvsOT}
\end{figure}

A more fine-grained analysis of the impact of $OT$ on detector performance is presented in terms of mAP and LAMR in Fig.~\ref{fig:mAPvsOT} and Fig.~\ref{fig:LAMRvsOT}. YOLOv3 delivers the best performance for gaze and hand labelled data in terms of both mAP and LAMR. In $S_G$, YOLO based object detectors deliver mAP greater than 50\% for $OT\leq 0.4$, whereas the Faster R-CNN delivers a mAP greater than 50\% for $OT < 0.2$.

%In terms of $precision$ and $recall$ ($sensitivity$), the values in Table~\ref{tab:Results} indicate that the gaze based detectors, in $S_G$, were comparatively more precise than the hand based detectors, in $S_H$, and produced a smaller number of false positives (relative to detectors in $S_H$). However, the recall for gaze based detectors was much lower. One explanation for this could be merger of neighbouring ROIs into a single ROI. Evidence of this can be found in Fig.~\ref{fig:Results} where we present four sample images containing predictions from the best performing object detectors in $S_G$ and $S_H$ which are the Faster R-CNN and the YOLOv4, respectively. It can be observed in row-3 of Fig.~\ref{fig:Results} that three closely spaced ROIs were detected as a single large ROI by the detector in $S_G$, whereas the detector in $S_H$ was able to detect them as three distinct ROIs. Even though the predicted, single ROI encompasses all three of the actual ROIs; while calculating the recall for the gaze detector, in the case presented, we considered the True Positive$(TP)=1$ (instead of $TP=3$) and False Negative$(FN)=2$ which resulted in a lower recall value overall. 

\subsubsection{Labelling Time}
\label{subsubSec:LabelTimeResults}
Two pathologists were asked to label 50 medium size KPs using each of the annotation technique and the time taken to label them all was measured using a stopwatch. We then determined the average time to label a single KP. The results of these experiments are presented in Table~\ref{tab:Time Taken} and discussed in the next section. Pathologist B consumed a larger amount of time for creating freehand labels as compared to pathologist A. The annotation time for both pathologists while generating bounding box based labels was similar. For gaze based labels, pathologist A was again slightly faster. The average of both annotators' times (Table~\ref{tab:Time Taken}) indicates that bounding box annotations consumed approximately 65\% less time than freehand annotation. Gaze annotations were the fastest and took less than half of the time consumed by bounding box annotations and 85\% less than the time for pixel wise annotations. With each time saving annotation methodology, we trade-off granularity and accuracy. Our goal was to examine the trade-off between granularity and time saving in gaze based annotations; the results in this section enable us to quantify this trade-off. We believe that labeling quality can be improved by developing more advanced gaze based annotation tools.

\subsection{Discussion}
\label{subSec:Discussion}
The experiments conducted for this work revealed a number of key insights about gaze based data annotation which we believe will be useful for anyone planning to explore this area in more detail. We now discuss our findings in the context of the questions posed in section~\ref{sec:Intro}.

\emph{Q-1:} The gaze annotation process does contain a fair bit of noise and comes with its own unique set of challenges and characteristics. The first is that eye gaze can be unintentional and influenced by the subconscious, making it more prone to errors and noise as compared to hand labeling where, gestures are deliberate and based on conscious decisions to annotate. It is also difficult for a human mind to stay focused on the task for long; so if the annotator zones out while inadvertently looking at a particular region or becomes distracted by something else on the screen, that region will get annotated as an ROI. Conversely, the annotator might already be looking at a candidate region while making the decision about its status as an ROI or not. As gaze recording is a continuous process, the area may get falsely annotated as an ROI. However, our results demonstrate that even though false ROI labels do occur, this is not a substantial problem. This is evident from the fact that YOLO based object detectors in $S_G$ were able to achieve performance that was close to the baseline detectors in $S_H$ (for $OT\leq 0.4$) despite there being no rigid constraints on the annotators during gaze labeling.
%We can make this claim because the performance degradation compared to hand labeling was not too large even though our experiments did not place any strict constraints on the annotators during gaze labeling. 
However, we still believe that improved software with user interfaces (UIs) designed specifically for gaze based annotations are  required and have the potential to match the performance of hand labeling. Integrating gaze labeling with other modalities such as voice commands or keystrokes may also provide performance gains by turning off gaze labeling when not required to avoid falsely labelled ROIs.

\emph{Q-2:} The performance of two out of the three detectors in $S_G$ is close to the their corresponding detectors in $S_H$ at lower values of $OT$ indicating that gaze based training does provide useful information. This seems to indicate that gaze based labeling even with simple preprocessing techniques like KDE enables estimation of location and (to some extent) size of objects. Furthermore, use of more advanced preprocessing approaches may deliver further improvements in performance.

\emph{Q-3:} In terms of labeling time, it seems that gaze labeling does indeed require less time as compared to conventional labeling techniques. Again, further improvements in UIs and tracking hardware are likely to yield further reduction in labeling time. We would like to highlight that the labeling time in Table~\ref{tab:Time Taken} currently does not include the initial time consumed for tracker calibration which was required only once at the beginning of a new labeling session and took around 5 to 10 seconds on average.

\emph{Q-4:} In this work our focus was primarily on ROIs with simple circular or oval shapes. Incorporating prior knowledge about the shape of target ROIs does seem to allow us to get rough estimates of the shape. For example, use of Gaussian kernels enabled us to model the rough shapes of KPs. However, we did not investigate this aspect in detail and our results at this stage are therefore, insufficient for us to be able to make concrete conclusions regarding shape estimation. 
%Our current results demonstrate that estimation of location and (to some extent) size of objects is possible with reasonable accuracy using gaze based labeling. 
A limitation of our method is manual selection of the kernel size $\sigma$ in WSIs containing ROIs of drastically different sizes. As mentioned in section~\ref{subSec:KDE}, when dealing with ROIs of different sizes, our current approach employs two different values of $\sigma$ on gaze data from a single recording session and then merges the two binary masks.

\begin{figure*}[t]
\centering
\includegraphics[width=\textwidth]{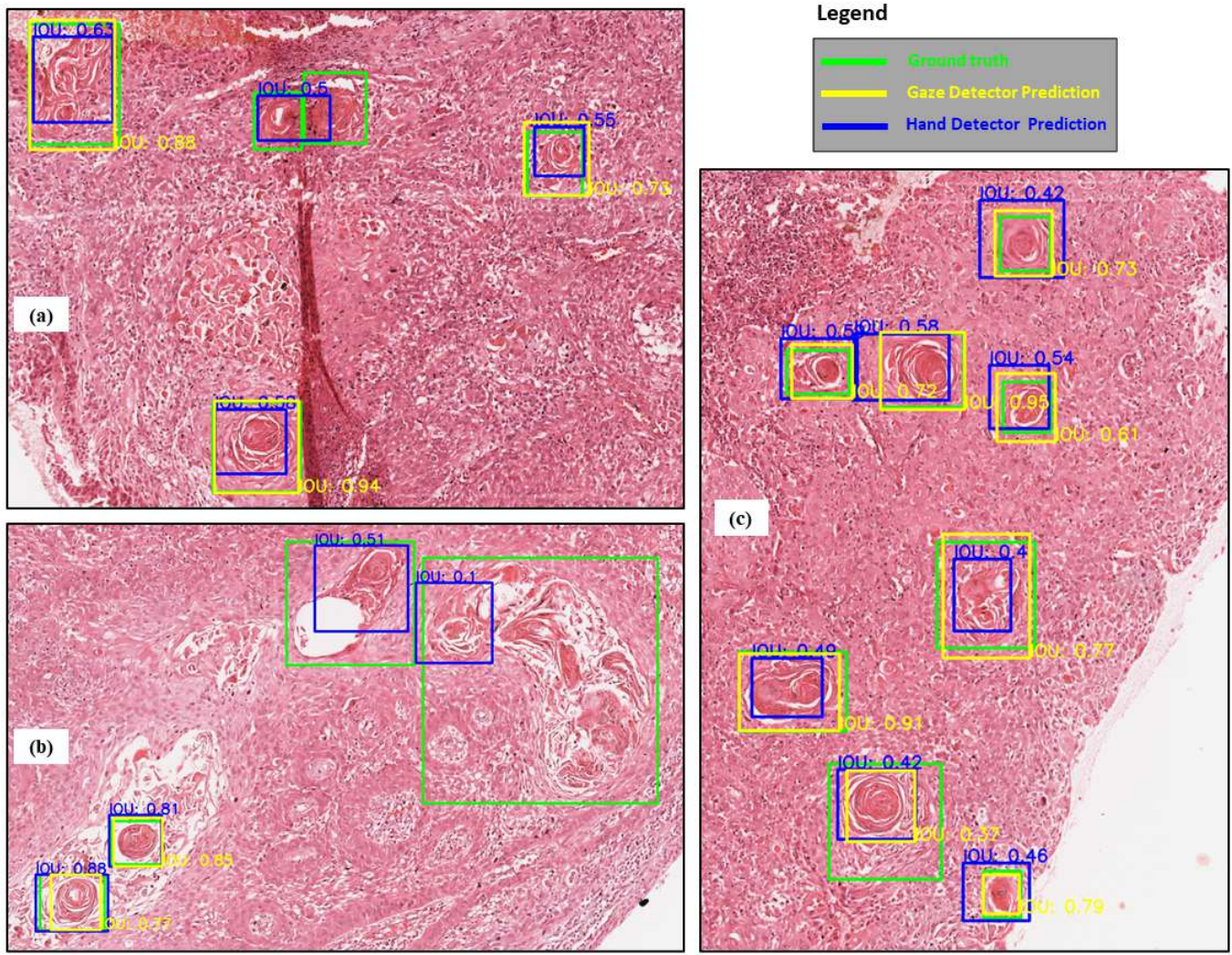}
\caption{Samples results of the YOLOv3 hand and gaze based detectors. Numerical values around a box indicate the IOU between a predicted bounding box and  its corresponding ground truth bounding box.}
 \label{fig:SampleResults}
\end{figure*}

\emph{Q-5:} Three well-established object detectors were tested using gaze and hand labelled data. Extensive performance evaluations were conducted to assess the difference in performance. Our analysis demonstrates that when trained using gaze labelled data, YOLO based object detectors demonstrate performance that is similar to that of hand labelled data at $OT\leq0.4$. The performance gap becomes greater at higher values of $OT$. Conventional object detection applications typically use $OT=0.5$. However, conventional object detection approaches such as pedestrian detection are required to deal with challenges such as occlusion of target objects behind other objects that occur when 3D scene information is projected into a 2D image. The nature of WSI data is inherently different and occlusion-free. Therefore, the use of high $OT$ values may be too restrictive in this scenario. Some sample results of the best performing detector (YOLOv3) are shown in Fig.~\ref{fig:SampleResults}; it can be observed that in practice, the gaze detector predictions are reasonably good. Finally, it is highlighted that noticeably better performance of the YOLOv3 detectors on gaze labelled data compared to other detectors (YOLOv5 and Faster R-CNN) seems to indicate that detector choice can also have an impact on performance on gaze data. Therefore, custom-built detectors may allow further improvements in performance on gaze labelled data.

\section{Conclusions and Future Work}
\label{sec:Conclusion}
We have presented the first of its kind work that thoroughly investigates the performance of gaze based labeling of histopathology WSI data.  We have demonstrated that use of gaze based labeling is sufficiently accurate to be able to train machine learning algorithms for object detection and classification. However, gaze-labelled object detection does have limitations which still need to be overcome. It also does not convey accurate shape information as accurately as `Freehand' hand labeling.
We believe that labeling quality can be enhanced further by making software improvements to the gaze labeling pipeline and/or employing better gaze tracking hardware. For the  KP segmentation problem exploiting prior knowledge of the ROI shape seems appealing. A potential approach can be purpose-built deep CNNs that search for oval/circular shaped ROIs.
These improvements may allow extension to more challenging applications like semantic segmentation. It would be interesting to see if performance gains can be obtained by employing ensemble based approaches similar to the ones proposed in \cite{menze2014multimodal, kavur2020basic, conze2021abdominal, kavur2020comparison}. However, we have left this for future work since the focus of this work is to examine the performance difference between hand and gaze based labeling and not performance maximization.
We have also shown that gaze recordings contain a number of outliers as a natural consequence of the pathologist searching or studying slides as they eye ball the WSI in search of ROIs. A majority of these points can be filtered out using simple pre-processing techniques.

We have also demonstrated that the time and effort required by pathologists can be conserved using gaze annotations as an alternative technique to conventional annotation strategies.
Our current pipeline requires manual selection of the threshold parameter. However, this can be overcome by developing adaptive thresholding approaches and we plan to tackle it as part of our future work. Even in its current state, the burden of the manual data clean up and post-processing lies with the data analysis team while saving the pathologists and domain experts valuable time that may be utilized for tasks other than data labeling. 

Our work also required the development of a unique software platform that can be interfaced with gaze tracking hardware for annotation of WSI data. This tool has the potential to be expanded to other biomedical image and video file formats for gaze recording thus providing a platform for other eye gaze related research. Overall we believe that gaze based labeling is currently in its infancy, it has tremendous potential but requires development of dedicated labeling software and UIs. As part of our future work, we will develop the gaze based equivalent of tools such as the ``Magic Wand'' in QuPath, that speed mouse based annotations of data. Another avenue worth exploring is whether or not to include a visual feedback mechanism for the pathologist? In our current setup when the pathologist focused on an ROI, no feedback was provided about the density/magnitude of gaze points on screen. Also, there was no feedback about whether or not a candidate region was successfully tagged as an ROI. Keeping gaze points invisible at the time of labeling was intentional since it became a source of distraction. The downside of these invisible data points was that it was somewhat challenging to get reassurance on whether or not labeling had been stored correctly. Hence it was possible for the pathologist to revisit a ROI and attempt to label it again. Selecting the right type and amount of feedback is therefore an avenue that requires further investigation in the future. 
Another interesting avenue of future research can be to conduct a thorough comparison of our constraint-free approach with constrained approaches such as \cite{Vilarino2007} which require the use of an On-Off switch by the Pathologist to trigger labeling of an ROI. Replacing the simple On-Off switch with a mono-stable switch may eliminate the need for the fixation time (required by our current setup) and the associated uncertainty about whether it has lasted long enough to be reliably registered. Also the false positive registrations may decrease with such a design. However, the false negatives may increase in case the Pathologist forgets to press the label trigger.

In summary, we conclude that gaze based labeling, with minimal constraints, saves time and effort and  is able to deliver labels that after pre-processing and noise removal are good enough to train machine learning algorithms on simple object detection and classification tasks. However, there is significant margin for improvement and performance is expected to improve with better tools and gaze tracking hardware. We hope that the data and tools being released with this work will pave the way for other researchers to delve deeper into this challenging research problem.
%%%%%%%%%%%%%%%%%%%%%%%%%%%%%%%%%%%%%%%%%%%%%%%%%%%%%%%%%%%%%%%%%%%%%%%%%%%%%%%%%%%%%%%%%%%%%%%%%%%%%%%%%%%%%%%%%%%%%%%%%%%%%%%%%%%%%%%%%%%%%%%%%%%%%%%%%%%%%%%%%%%%%%%%%%%%%%%%%%%%%%%%%

%%%% The lines below are for color journal look etc %%%%
%\smallskip\noindent
%\begin{small}
%\begin{tabular}{l}
%\verb+\+\texttt{documentclass[journal,twoside,web]\{ieeecolor\}}\\
%\verb+\+\texttt{usepackage\{\textit{Journal\_Name}\}}
%\end{tabular}
%\end{small}

\bibliographystyle{./bibliography/IEEEtran}
% Generated by IEEEtran.bst, version: 1.12 (2007/01/11)

\end{document}